# Noisy Quadrature of Squeezed Light and Laser Cooling

G.M. Saxena and  A Agarwal

National Physical Laboratory

Dr. K.S.Krishan Road, New Delhi 110012, India

e-mail: gmsaxena@mail.nplindia.ernet.in

**Abstract:**

The laser cooling of atoms is a result of the combined effect of Doppler shift, light shift and polarization gradient. These are basically undesirable phenomena. However, they combine gainfully in realizing laser cooling and trapping of the atoms. In this paper we discuss the laser cooling of atoms in the presence of the squeezed light with the decay of atomic dipole moment into noisy quadrature. We show that the higher decay rate of the atomic dipole moment into the noisy quadrature, which is also an undesirable effect, may contribute in realizing larger cooling force vis-à-vis normal laser light.

**Introduction:**

The laser cooling of the atoms and ions has been a subject of extensive studies in recent years [1]. The laser cooling of atoms has number of applications i.e., in the Cs fountain clock, Bose-Einstein condensation, test of principles of Quantum mechanics and the study of time dependence of the fundamental constants. In the laser-cooling experiment, the polarization gradient [1] and dark state [2] techniques are used for cooling the multi-level atoms to the temperatures much lower than the 'Doppler limit'. However, very low temperatures may also be reached by using the squeezed light [3] even for the two-level atoms. It has been shown [4] that the temperatures slightly below the Doppler limit can be obtained for the two-level atoms in a near-resonant standing wave squeezed field. We establish in this paper that by using the noisy-field quadrature of the squeezed vacuum under certain phase matching conditions, temperature much lower than the Doppler limit may be reached. We shall see that the decay of the atomic dipole moment into the noisy quadrature, an undesirable feature of the squeezed light, results in higher cooling force. It is quite understandable as the laser cooling is after all the result of combined effect of the undesirable phenomena like light shift, Doppler shift and polarization gradient. How well these undesirable effects have helped in cooling the atoms to very low temperature is a well-established observation. And now we have one more undesirable feature that is the noisy quadrature of the squeezed vacuum. The atoms in the noisy quadrature decay at a rate faster compared to the spontaneous emission rate. We report in this paper that the noisy quadrature of the squeezed light can be gainfully used in realizing the higher cooling force vis-a-vis normal laser light.

In this paper we study the cooling force on the Cs atoms in the optical molasses formed by the squeezed vacuum and squeezed coherent states with certain relative phase matching conditions. The slow moving Cesium atoms are cooled and trapped in this optical molasses. The squeezed vacuum with proper phase interacts with those atoms, which have enhanced dipole decay rate in the noisy quadrature. We show that for the combination of squeezed vacuum and squeezed-coherent light, the atoms can be cooled much below the Doppler limit even for a resonant driving field. It occurs when the phase of the squeezed vacuum relative to driving laser field is 0 or $\pi$ and the phase of the squeezed coherent light is different from 0 or $\pi$.

**Squeezed Light Cooling Force:**

In this section we shall discuss the cooling force on the atoms with the squeezed light. We first, briefly, describe one of the methods of generating the squeezed light using an optical parametric oscillator (OPO) [5] before considering the cooling of atoms in the presence of the cw source of the squeezed light. The description of the squeezed light generation using OPO is included for defining the parameters that characterize the squeezed states. In the set-up for the squeezed light an external-cavity diode laser locked to Cs transition at 852 nm serves as a primary source of light or pump for the input of OPO. The laser light at 852 nm is first sent to a frequency-doubling cavity. The cavity output produces a pump light at 426 nm for OPO and it is frequency locked to the incident laser. The doubling cavity contains a Potassium Niobate ($KNbO_3$) nonlinear crystal. The OPO cavity is a four-mirror folded cavity with two curved and two plain mirrors. The non-linear medium at the waist between curved mirrors is an a-cut $KNbO_3$ crystal. The spontaneous parametric fluorescence, produced by degenerate down conversion into a sub-harmonic mode of OPO, generates the squeezed vacuum. In a degenerate OPO, the pump field at $2\omega$ is split by a non-linear crystal into two photons of frequencies $\omega$ each. The average values of the photon annihilation and creation operators in the squeezed vacuum are $\langle a_k \rangle = \langle a_k^+ \rangle = 0$ [6] and that of second order photon annihilation and creation operators are $\langle a_k^+ a_k \rangle = N(\omega)$, $\langle a_k a_k \rangle = M(\omega)e^{i\phi}$, here $N(\omega)$ corresponds to the photon number, $M(\omega)$ is the two-photon correlation function and $\phi$ is the squeezed field phase. The photon number and two-photon correlation functions are not independent of each other. They satisfy the inequality $M^2(\omega) \leq N(\omega)[N(\omega) + 1]$ in the quantum picture [7]. The factor (N+1) instead of N arises due to the quantum nature of the field. For an ideal degenerate OPO [8], we have

$N(\omega) = (\lambda^2-\mu^2)[1/(\omega^2+\mu^2) - 1/(\omega^2-\lambda^2)]/4$,

$M(\omega) = (\lambda^2-\mu^2)[1/(\omega^2+\mu^2) + 1/(\omega^2-\lambda^2)]/4$ and $M(\omega)^2 = N(\omega)[N(\omega) + 1]$, where $\lambda = \kappa/2 + \varepsilon$, $\mu = \kappa/2 - \varepsilon$, with $\kappa$ and $\varepsilon$ being the cavity decay constant and the amplification factor respectively for OPO. The values of N and M functions may be selected by the OPO parameters. The OPO generates an ideal squeezed vacuum with $N = \sinh^2 r$, and $|M| = \cosh r \sinh r$, here r is the squeeze factor. The signal and idler outputs of the OPO may be combined at a 50/50 beam-splitter to produce a squeezed-vacuum state and a squeezed-coherent state [9].

In the conventional laser-cooling set-up for the fountain atomic clock, the Cs atoms are cooled and trapped by an optical molasses formed by three mutually perpendicular standing-wave laser beams. The cooling force due to the optical molasses has been calculated by several groups and is well-documented [10]. The cooling force on the slowly moving atoms with the density matrix $\rho$, in a standing-wave squeezed-coherent field [11,12] is given by

$F = i\hbar q_r (\Omega^*\langle\rho_{12}\rangle - \Omega\langle\rho_{21}\rangle) = \hbar q_r \Omega \langle\sigma_Y\rangle_{sc}$,  (1)

here $q_r = -k \tan(\mathbf{k} \cdot \mathbf{x})$, $k = 2\pi/\lambda$, $\lambda = 852$ nm, and $\Omega$ is the Rabi frequency of the optical field. $\langle\sigma_Y\rangle_{sc}$, the y-component of the atomic Block vector in the presence of the squeezed-coherent light is given by

$\langle\sigma_Y\rangle_{sc} = \Omega(\delta + \gamma|M|\sin\phi)/2D_{sc}$,  (2)

where δ is the laser frequency detuning from the resonance, γ is the normal atomic decay rate, φ is the relative phase of the driving laser field with respect to the squeezed-coherent light, and

$$D_{sc} = n\,(\tfrac{1}{4}\gamma^2 n^2 + \delta^2 - \gamma^2 |M|^2) + \Omega^2\,(n/2 + |M|\cos\phi), \qquad (3)$$

and n = 1 + 2N. The force given by (1) is calculated assuming **k.v** < γ and δ < γ, where **v** is the velocity of the atoms. We observe from (1) and (2) that the magnitude of the cooling force basically depends on (δ + γ|M| sin φ) and there is non-vanishing cooling force even at zero detuning (δ=0). The term γ|M| sin φ in the expression for cooling force arises due to the interaction of atoms with squeezed-coherent light- a quantum field. From this term it follows that the squeezed light enhances the cooling force. The two-photon correlation factor M depends on the degree of squeezing. The decay rate γ is also an important parameter in determining the cooling force. Several authors have shown that the squeezed states might be used for manipulating the atomic spontaneous decay rate [8,9]. For increased decay rate a larger cooling force may be realized. On interacting with the squeezed vacuum, the atoms de-excite into either of the two quadratures with different decay rates. The atoms in the squeezed (in-phase) quadrature relax at a slower rate while those in the noisy (out-of-phase) quadrature decay at faster relaxation rate compared to normal vacuum. Specifically, when the relative phase Φ between the coherent driving field and the squeezed vacuum is 0 or π, the atoms decay in the out-of-phase or the noisy quadrature with the larger decay rate $\gamma_x = \gamma(N + \tfrac{1}{2} + M)$. In general, the squeezed states are used for reducing the quantum noise or the decay rate of the atoms. However, in this paper, we demonstrate that even the noisy quadrature could be gainfully utilized in realizing higher cooling force than that with either laser light or quieter quadrature of the squeezed light.

Let us consider that the atoms in the optical molasses formed by the squeezed-coherent light, decay with the faster rate $\gamma_x$ into the noisy quadrature of the squeezed vacuum of phase $\phi$. We calculate the cooling force on the atoms in the combined presence of the squeezed vacuum and squeezed coherent states. On incorporating the faster decay rate in (1), the cooling force is,

$$F_{sv} = q_r \Omega \langle \sigma_Y \rangle_{svsc}, \qquad (4)$$

Here $\langle \sigma_Y \rangle_{svsc}$, the y-component of the atomic Bloch vector in the combined presence of the squeezed vacuum and squeezed coherent states, is given by the expression

$$\langle \sigma_Y \rangle_{svsc} = {}^1\!/_2\, \beta\, [\Delta + 2\,|M|\,(n/2 + |M|)\sin\phi]\, /\, D\,, \qquad (5)$$

$$D = n\,/\,_2\, [\beta^2 + 2\Delta^2 + (n/2 + |M|)^2] + \beta^2\,|M|\cos\phi\,. \qquad (6)$$

The normalized detuning $\Delta$ and Rabi frequency $\beta$ are defined as $\Delta \equiv \delta/\gamma$ and $\beta \equiv \Omega/\gamma$. It may be readily shown that with the modified y-component of the atomic Block vector $\langle \sigma_Y \rangle_{svsc}$, the cooling force is increased depending on the phase $\phi$ and squeeze parameter r. The cooling force is calculated for the different values of these parameters.

The relevant parameter quantifying the degree of squeezing is given by the factor 2(|M| - N). Georgiades, et. al [13] have developed a source of squeezed light exhibiting approximately 75% squeezing. This corresponds to N = 9/16, |M| = 15/16 and r ≈ 0.7. With this kind of source of squeezed light one can efficiently cool and trap atoms. We have calculated the cooling force for squeezed light and for the combination of squeezed vacuum and squeezed coherent light, and denoted them as F and $F_{SV}$ respectively. We compare the cooling force F and $F_{sv}$ in figures 1 and 2 for different cases. As shown in fig. 1, for small values of Rabi frequency $\beta$, F (shown by dashed line) is greater than $F_{sv}$ (shown by solid curve). For $\beta > 3$, this trend reverses. There is a steep rise in $F_{sv}$ compared to F for the increasing value of $\beta$. Figure 1(a) is a plot for the resonant case. We find that the squeezed light can give rise to a cooling force even at zero detuning. In fig 1(b), we plot the off-resonance case. This shows similar behavior as in Fig. 1(a) except that the cooling force

increases slightly. For squeezed (in-phase) quadrature instead of the noisy quadrature, i.e., for the phase $\Phi = \pi/2$, the cooling force (shown by dotted line in fig. 1) is much smaller than F (shown by dashed line). This shows the superiority of the noisy quadrature to the in-phase quadrature of the squeezed light for efficient cooling of the atoms.

In fig.2, we show that the relative phase $\phi$ and the degree of squeezing play significant role in lowering the temperature in the noisy quadrature of the squeezed light. Figures 2(a) and 2(b) are for the resonant and off-resonant squeezed light respectively. These figures show that the cooling force is sensitive to small changes in the relative phase between fields, the detuning from the resonance, value of Rabi frequency and degree of squeezing. The atoms in the presence of the squeezed vacuum, when its phase $\Phi$ relative to the laser field is 0 or $\pi$, can attain minimum temperature much lower than that given by Doppler limit $k_B T_D = h\gamma/2$ here $k_B$ is the Boltzman constant.

**Conclusion**:

We have established in this paper that the noisy quadrature could be gainfully utilized in realizing higher cooling force than that with either laser light or with the squeezed vacuum. Besides, for the certain values of the phase of the squeezed coherent field with respect to the laser field and for the appropriate value of two-photon correlation M, atoms can be cooled even for the resonant squeezed light. It is very unexpected result because the noisy quadrature is invariably treated as undesirable feature of the squeezed light.

**References:**

[1]    D.Wineland and W.Itano, Phys.Rev. A 20, 1979, 1521; S.Chu and C.Wieman, J. Opt Soc. Am B 6, 1989,1961; J.Dalibard and C. Cohen-Tannoudji, J. Opt. Soc. Am B 6, 1989, 2023.

[2]    A. Aspect, E. Arimondo, R. Kaiser, N. Vansteenkiste and C.Cohen-Tannoudji, J. Opt. Soc. Am B **6** (1989) 2112.

[3]    Y. Shivey, Phys. Rev. Lett. **64** (1990) 2905.

[4]    R Graham, D F Walls, and W P Zhang, Phys. Rev. A **44** (1991) 7777.

[5]    E. S. Polzik, J. Carri and H.J. Kimble, Appl. Phys B **55** (1992) 279.

[6]    C.W.Gardiner,Phys. Rev. Lett. **56** (1986) 1917, A.S. Parkins and C.W.Gardiner, Phys. Rev A **40** (1990) 3796.

[7]    H.J. Carmichael, A.S. Lane and D.F.Walls, Phys. Rev. Lett. **58** (1997) 2539; *ibid*. J. Mod. Opt. **34** (1987) 821.

[8]    H. Ritsh and P. Zoller, Opt. Comm. **64** (1987) 523; *ibid*. Phys. Rev. A.**38** (1988) 4657.

[9]    C. Kim and P. Kumar, Phys. Rev. Lett. **73** (1994) 1605.

[10]   H. J. Metcalf and P. van der Straten, "Laser Cooling and Trapping", Springer Verlag New York, 1999.

[11]   B. J. Dalton, Z. Ficek and S Swain, J. Mod. Opt. **46** (1999) 379.

[12]   Z. Ficek, W.S.Smyth and S. Swain, Phys. Rev. A **52** (1995) 4126.

[13]   N.P. Georgiades, E.S. Polzik, K. Edamatsu, H.J.Kimble and A.S.Parkins, Phys. Rev.Lett. **75** (1995) 3426.

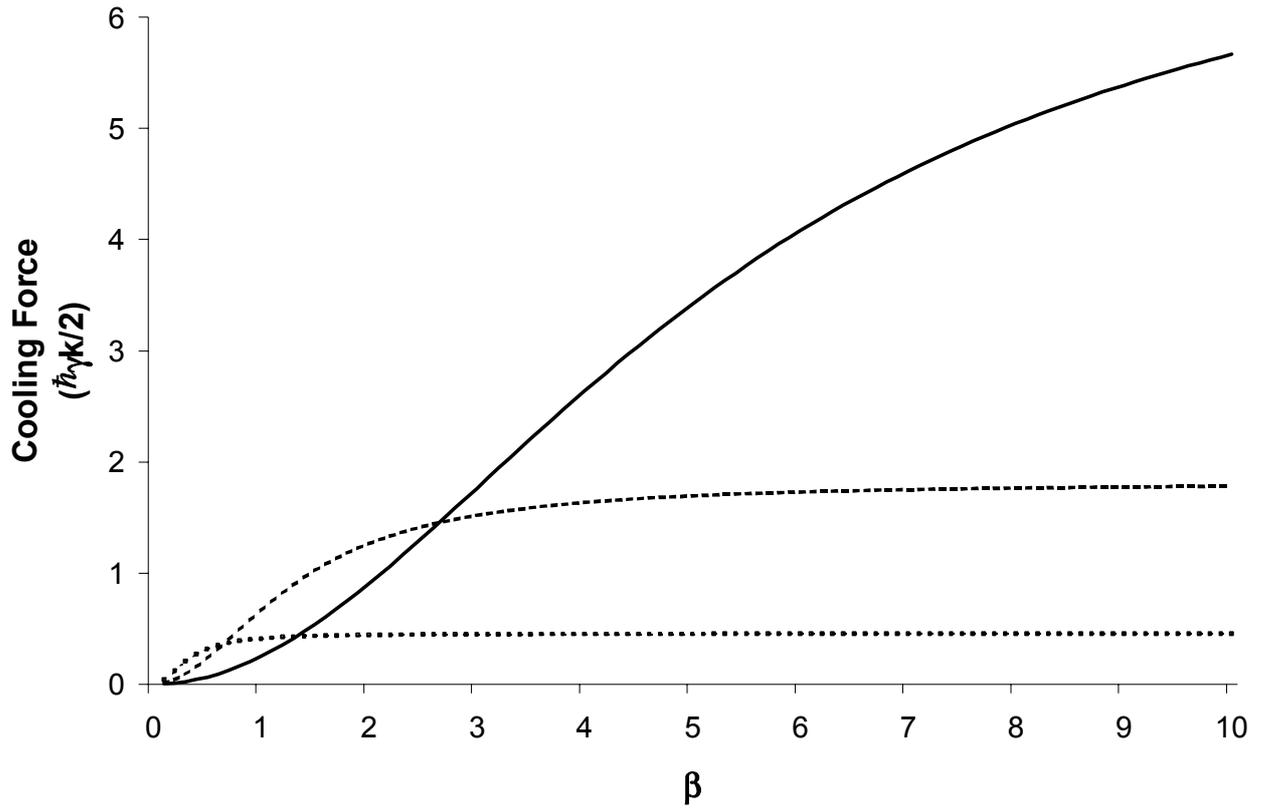

Fig.1(a) The dependence of spatially averaged force on the Rabi frequency $\beta$ of the driving field in the presence (solid and dotted line) and absence (dashed line) of squeezed vacuum. The solid line is for $\Phi = 0$ or $\pi$ (out-of-the phase) and dotted line is for $\Phi = \pi/2$ (in-phase quadrature). Here degree of squeezing is chosen to be 75% and $\phi = 0.8\,\pi$. for the resonant case $\Delta = 0$

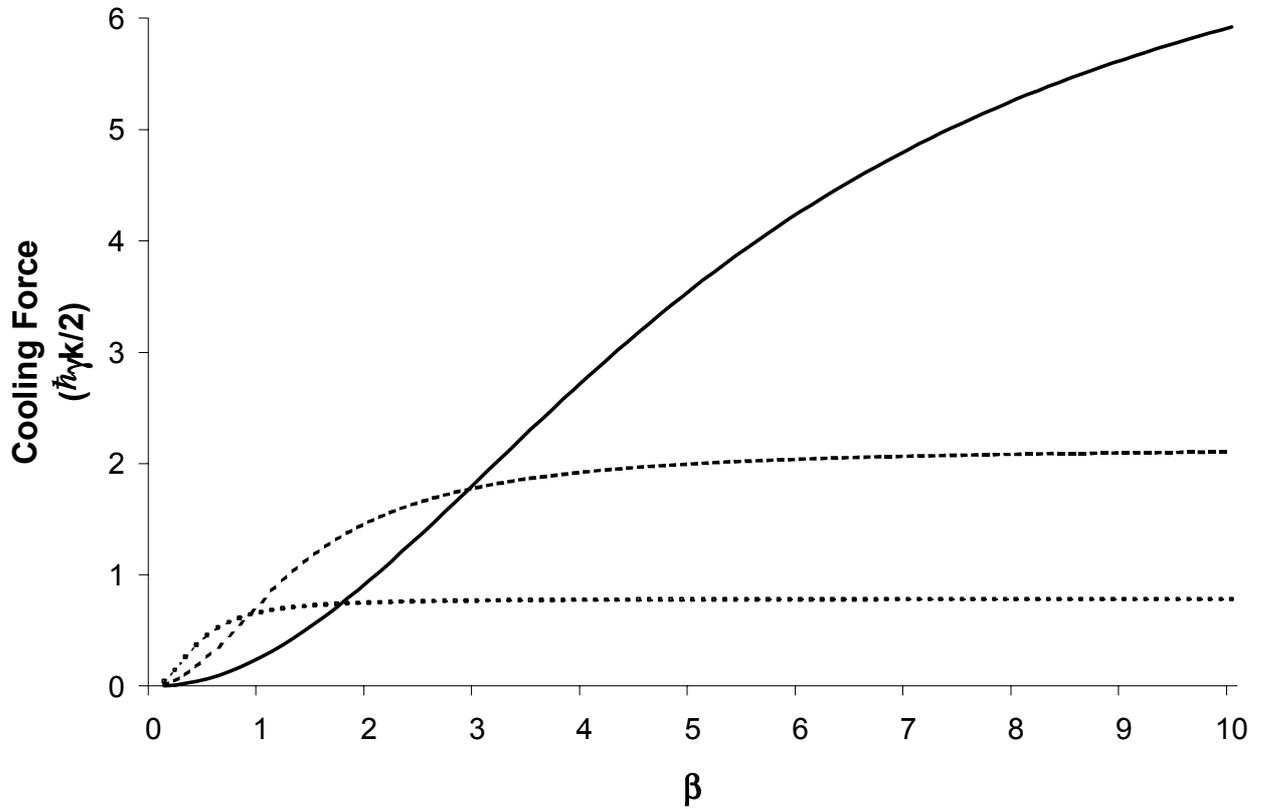

Fig-1(b) The dependence of spatially averaged force on the Rabi frequency β of the driving field in the presence (solid and dotted line) and absence (dashed line) of squeezed vacuum. The solid line is for Φ = 0 or π (out-of-the phase) and dotted line is for Φ = π/2 (in-phase quadrature). Here degree of squeezing is chosen to be 75% and ϕ = 0.8 π. for the off-resonant case Δ = 0.1. Cooling Force hγk/2 = 1 is the Doppler Limit.

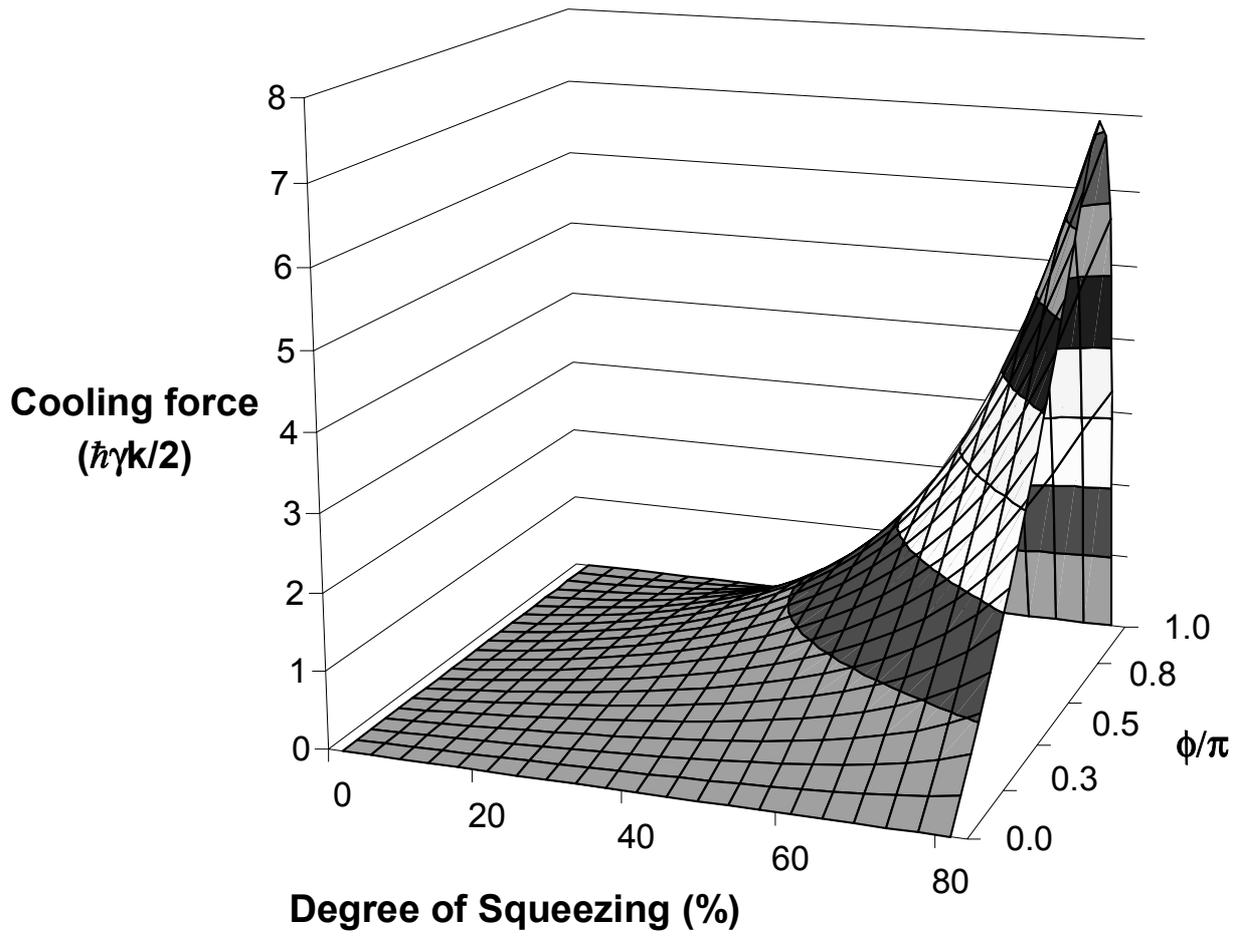

Fig.2(a) A three-dimensional plot of spatially averaged cooling force $F_{sv}$ as a function of the degree of squeezing and relative phase $\phi$ for the resonant case $\Delta = 0$.

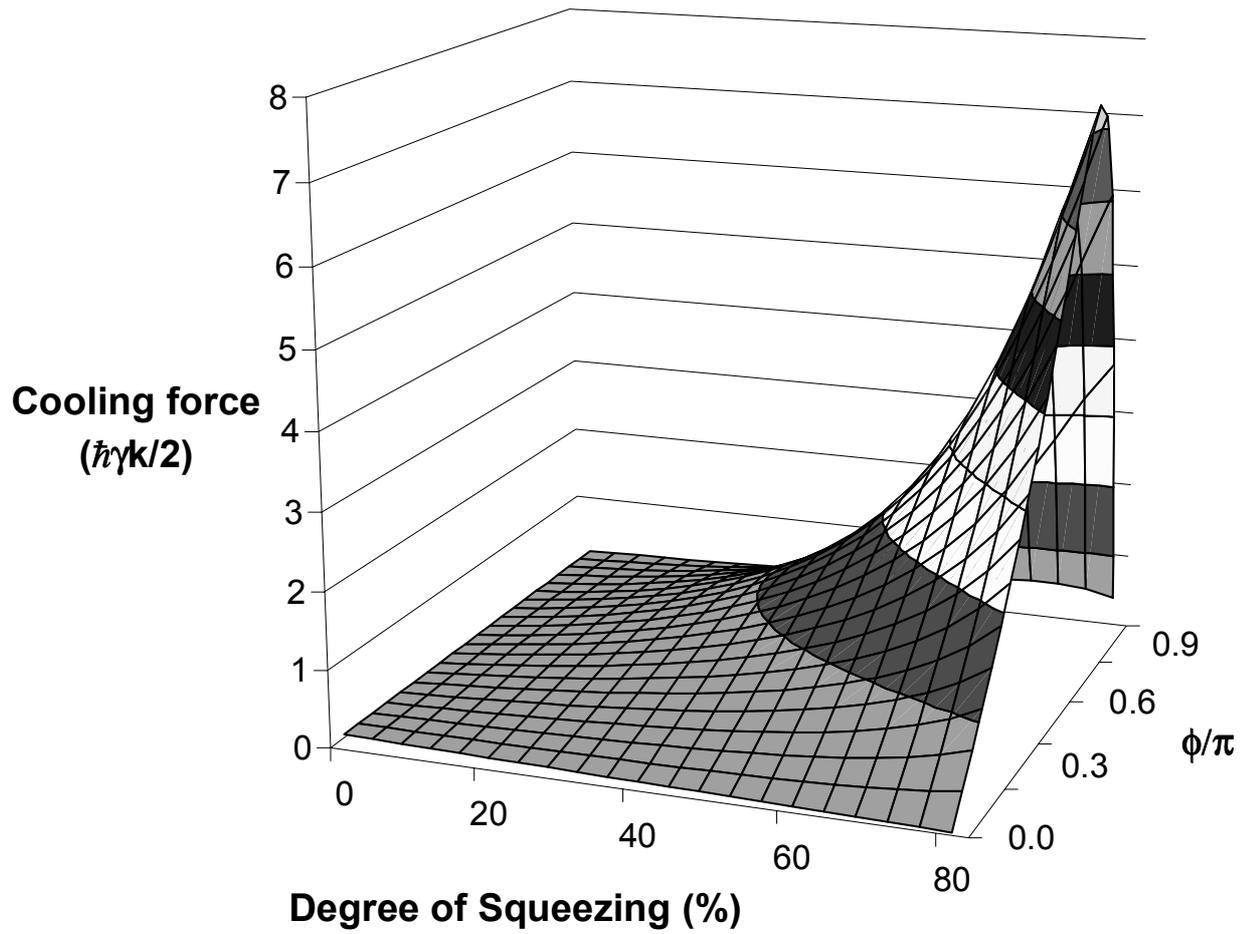

Fig. 2(b) a three-dimensional plot of spatially averaged cooling force $F_{sv}$ as a function of the degree of squeezing and relative phase $\phi$ for the off-resonant case $\Delta = 0.1$. Rabi frequency $\beta = 10$.